# Direct Observation of a Fractional Charge

R. de-Picciotto, M. Reznikov, M. Heiblum, V. Umansky, G. Bunin and D. Mahalu

*Braun Center for Submicron Research, Department of Condensed Matter Physics, Weizmann Institute of Science Rehovot 76100, Israel*

**Ever since Milliken's[1] famous experiment it is well known that the electrical charge is quantized in units of the electronic charge - *e*. For that reason, Laughlin's[2] theoretical prediction of the existence of fractionally charged *quasi particles*, put forward in order to explain the *Fractional Quantum Hall* (FQH) effect, is very counter intuitive. The FQH effect is a phenomenon that occurs in *a Two Dimensional Electron Gas* (2DEG) subjected to a strong perpendicular magnetic field. This effect results from the strong interaction among the electrons and consequently current is carried by the above mentioned *quasi particles*. We directly observed this elusive fractional charge by utilizing a measurement of *quantum shot noise*. Quantum shot noise results from the discreteness of the current carrying charges and thus is proportional to their charge, $Q$, and to the average current $I$, namely, $S_i=2QI$. Our *quantum shot noise* measurements unambiguously show that current in a 2DEG in the FQH regime, at a *fractional filling factor $\nu=1/3$*, is carried by fractional charge portions $e/3$; in agreement with Laughlin's prediction.**

The energy spectrum of a *Two Dimensional Electron Gas* (2DEG) subjected to a strong perpendicular magnetic field, $B$, consists of highly degenerate Landau levels with a degeneracy per unit area $p=B/\phi_0$, with $\phi_0=h/e$ the flux quantum ($h$ being Plank's constant). Whenever the magnetic field is such that an integer number $\nu$ (the *filling factor*) of Landau levels are occupied, that is $\nu=n_s/p$ equals an integer ($n_s$ being the 2DEG areal density), the longitudinal conductivity of the 2DEG vanishes while the Hall conductivity equals $\nu e^2/h$ with very high accuracy. This phenomenon is known as the Integer Quantum Hall (IQH) effect[3]. A similar phenomenon occurs at **fractional** filling factors, namely, when the *filling factor* equals a rational fraction with an odd denominator $q$ and is known as the Fractional Quantum Hall effect[4]. In contrast to the IQH effect, which is well understood in terms of non interacting electrons, the FQH effect can not be explained in such terms and is believed to result from interactions among the electrons, brought about by the strong magnetic field.

Laughlin[2] had argued that the FQH effect could be explained in terms *quasi particles* of a fractional charge - $Q=e/q$. Although his theory is consistent with a considerable amount of the experimental data, no experiment directly showing the existence of the fractional charge exists. The early Aharonov-Bohm measurements[5] were proved to be in principle inadequate to reveal the fractional charge[6]. A more recent experiment based on resonant tunneling by Goldman and Su[7] was reproduced and interpreted differently by Franklin *et al*[8]. The difficulty in such experiments is that the results provide only the average charge per state and not the charge of individual particles. *Quantum shot noise,* on the other hand, probes the temporal behavior of the current and thus offers a direct way to measure the charge. Indeed, as early as in 1987, Tsui[9] suggested that the *quasi particle*'s charge could in principle be determined by measuring *quantum shot noise* in the FQH regime. However, no theory was available until Wen[10] recognized that transport in the FQH regime could be treated within a framework of One Dimensional (1D) interacting electrons, propagating along the edge of the two dimensional plane, making use of the so called Luttinger liquid model. Based on this model subsequent theoretical works[11] predicted that *quantum shot noise*, generated due to weak backscattering of the current, at fractional filling factors $\nu=1/q$ and at zero temperature, should be proportional to the *quasi particle*'s charge $Q=e/q$ and to the backscattered current $I_B$:

$$S_i = 2QI_B. \quad (1)$$

In order to realize such a measurement we utilized a Quantum Point Contact (QPC) - a constriction in the plane of a 2DEG - that partially reflects the current. The high quality 2DEG, embedded in a GaAs-AlGaAs heterostructure, some 100 nm beneath the surface, has carrier density $n_s=1\cdot10^{11}$ cm$^{-2}$ and mobility $\mu=4.2\cdot10^6$ cm$^2$/Vs at 1.5 K. The QPC is formed by two metallic gates evaporated on the surface of the structure, separated by an opening of some 300 nm which is a few Fermi wavelengths wide (see inset in Fig. 1). By applying negative voltage to the gates with respect to the 2DEG, thus imposing a local repulsive potential in the plane of the 2DEG, one may controllably reflect the incoming current. The sample was inserted into a dilution refrigerator with a base temperature of about 50 mK. Noise measurements where done by employing an extremely low noise home made preamplifier, placed in a 4.2 K reservoir. The preamplifier is manufactured from GaAs transistors, grown in our Molecular Beam Epitaxy system. The preamplifier has a voltage noise as low as 2.5 $10^{-19}$ V$^2$/Hz and a current noise of $1.1\cdot10^{-28}$ Amp$^2$/Hz at 4 MHz.

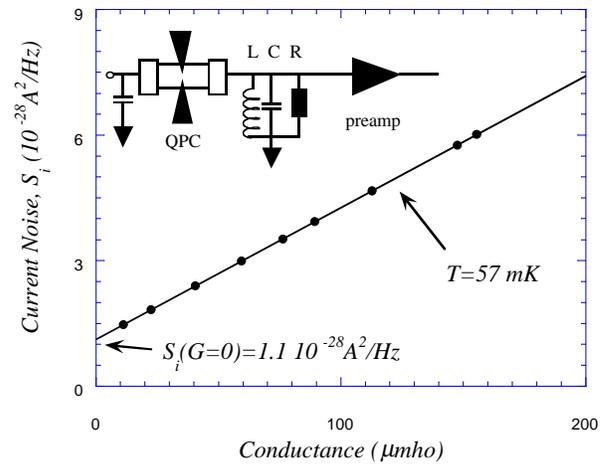

Fig. 1. The total current noise inferred to the input of the preamplifier as a function of the input conductance at equilibrium (full circles). The measured noise is a sum of thermal noise, $4K_BTG$ (leading to a straight line) and the constant noise of the amplifier. This measurement allows the determination of the temperature of the 2DEG. Inset: The quantum point contact in the two dimensional electron gas is shown to be connected to an *LCR* circuit at the input of a cryogenic preamplifier.

Current fluctuations, generated in the QPC, were fed into an *LCR* resonant circuit, with most of the capacitance contributed by the coaxial cable which connects the sample at 50 mK to the preamplifier at 4.2 K. Outside the cryostat the amplified signal was fed into an additional amplifier and from there to a spectrum analyzer which measured the current fluctuations within a band of ~100 KHz about a central frequency of ~4 MHz. Since the absolute magnitude of the noise signal is of utmost importance, a careful calibration of the total gain, from QPC to the spectrum analyzer, was done by utilizing a calibrated current noise source. This allows the *translation* of the spectrum analyzer output into spectral density of current fluctuations (current noise). Although our amplifier has excellent characteristics it still introduces current fluctuations into the circuit. This unwanted current noise must be subtracted from the total measured noise in order to extract the shot noise associated solely with the QPC. By measuring the total current noise while varying the conductance, $G$, of the unbiased sample (see Fig. 1), we deduce both the electron temperature, $T=(\delta S_i/\delta G)/4k_B$ and the contribution of our amplifier to the total noise (extracted from the extrapolated total noise to zero conductance). Note that the temperature we find, 57 mK, is very close to that of the sample holder.

Since the temperature, $T$, and the applied voltage, $V$, across the QPC during our measurement are both finite, the results must be compared with a more elaborate theory than the one leading to Eq. 1. The general expression for the *zero frequency* spectral density of the *total noise* in a single 1D channel is[12]:

$$S_i = 2g_0 t(1-t)\left[QV\coth\left(\frac{QV}{2k_BT}\right) - 2k_BT\right] + 4k_BTg_0t, \qquad (2)$$

where the transmission of the QPC, $t$, is given by the ratio between the conductance, $G$, and the quantum conductance, $g_0 = e^2/h$. This expression interpolates smoothly between the well known thermal noise $4k_BTG$ at equilibrium ($V=0$), and a linear current dependence at voltages much larger than $k_BT/Q$, for which $S_i \cong 2IQ(1-t)$, with $I=GV$. This dependence was experimentally verified for non interacting electrons[13] with $Q=e$. In the absence of a general theory for the FQH regime we are going to assume that Eq. 2 is valid for weak backscattering of quasi particles in the FQH regime at $\nu=1/3$, with $Q=e/3$ and $g_0 = e^2/3h$. Since this expression recovers both the equilibrium thermal noise for $V=0$ and $T\neq0$, and reduces to Eq. 1 for $T=0$ and $t\sim1$ ($Vg_0 t(1-t) = I_B \cdot t \approx I_B$), we believe that a correct theory should not deviate significantly from this expression..

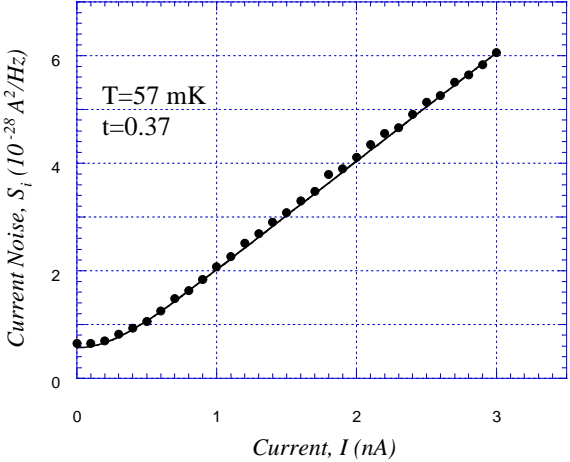

Fig. 2. *Quantum Shot Noise* as a function of DC current, $I$, through the QPC without an applied magnetic field (full circles). The solid line is Eq. 2 with the temperature deduced from Fig. 1.

*Quantum Shot noise* measurements as a function of the current through a partially pinched QPC were performed first in the absence of a magnetic field. The results, after calibration and subtraction of amplifier noise, are shown in Fig. 2. The transmission of the lower laying quasi-1D channel in the QPC is simply deduced from the measured conductance normalized by $2e^2/h$ (the factor 2 accounts for spin degeneracy). Our data fits almost perfectly the expected noise of Eq. 2 using the measured electron temperature without any fitting parameters.

The magnetic field was then swept from zero to 14 Tesla. The two-terminal conductance exhibits Hall plateaus, expected in the IQH and in the FQH regimes ($\nu=2/5, 3/5, 2/3$ and $1/3$ are clearly visible with a plateau width of ~1 Tesla around $\nu=1/3$). At a filling factor $\nu=1/3$ and full transmission (zero gate voltage) no excess noise above the thermal one is observed upon driving a current through the sample, thus ruling out noise which is related to overheating. The noise measured upon partially reflecting the current is drastically suppressed compared to the noise measured in the absence of a magnetic field as shown in Fig. 3.

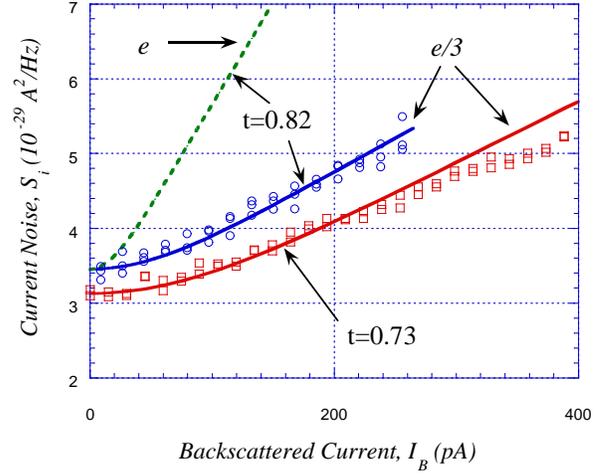

Fig. 3. *Quantum Shot Noise* as a function of the backscattered current, $I_B$, in the fractional quantum Hall regime at $\nu=1/3$ for two different transmission coefficients through the quantum point contact (open circles and squares). The solid lines correspond to Eq. 2 with a charge $Q=e/3$ and the appropriate $t$. For comparison the expected behavior of the noise for $Q=e$ and $t=0.82$ is shown by the dotted line.

Our data fit very well the expected noise of a current carried by *quasi particles* of charge $Q=e/3$. The backscattered current is calculated using the transmission, $t$, deduced from the ratio of the conductance to $g_0 = e^2/3h$. The slope of the noise versus backscattered current curve increases with applied voltage approaching the expected slope of $2te/3$ at voltages larger than $2k_BT/Q$ as expected. For comparison, the expected noise for $Q=e$ and the same $g_0$ is also shown.

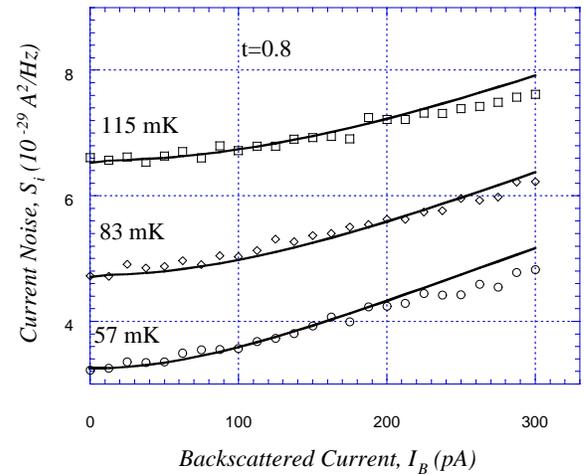

Fig. 4. *Quantum Shot Noise* as a function of backscattered current, $I_B$, in the fractional quantum Hall regime at $\nu=1/3$, for three different temperatures and a constant transmission coefficient, $t=0.8$, through the QPC.

The noise tends to saturate at even larger backscattered currents (note the deviation of the data points from the solid line). This additional noise suppression is accompanied by an onset of nonlinearity in the *I-V* characteristics (not shown). The nonlinearity in the FQH regime may result from the interaction among the electrons, an energy dependence of the bare transmission coefficient and from a finite excitation gap (a gap[9], $\Delta\approx250$ $\mu eV$, is expected at ~13 Tesla). These three sources are practically indistinguishable. Nonlinearity complicates the otherwise straightforward interpretation of our results and we thus choose to show data in a smaller voltage range and for moderate reflection coefficients where the *I-V* is linear.

In order to further investigate the behavior of quantum shot noise at the FQH regime, we measured the noise versus backscattered current for three different temperatures and a fixed transmission through the QPC (shown in Fig. 4). The data fit the curves expected from Eq. 2 with *Q=e/3*. Note that Eq. 2 with a charge *Q=e/3* suggests not only that the amplitude of the noise is proportional to *Q* but also that shot noise is observed above the thermal noise at a characteristic voltage *V=6k$_B$T/e,* three fold larger than the value for noninteracting electrons. This is because the potential energy of the quasiparticles is eV/3. The agreement between the data and the detailed shape of Eq.2 at small backscattered currents, gives thus an additional indication for the existence of a smaller charge e/3.

In conclusion, our noise measurements unambiguously show that the current in the FQH regime, at filling factor 1/3, is carried by *quasiparticles* with charge *e/3*. In contrast to conductance measurements, that measure an averaged charge over quantum states or over time, our *quantum shot noise* measurement is sensitive to the charge itself. The *"magic"* of an apparent smaller charge due to electron – electron interactions is a beautiful manifestation of the strength of the theoretical methods[2] used to predict such a counter intuitive behavior.

During the writing of this manuscript we became aware of similar work[14] in which the authors measure the same charge at a filling factor *ν=2/3,* also using shot noise measurements.

### Acknowledgments

The work was partly supported by a grant from the Israeli Science Foundation and by a grant from the Austrian Ministry of Science, Research and art (sektion Forschung).